\documentclass[aps,reprint,prb]{revtex4-1}
\usepackage{amsmath}
\usepackage{amssymb}
\usepackage{graphicx}
\usepackage{hyperref}
\begin{document}
\title{Pressure-induced ferroelectric phase of LaMoN$_3$}
\author{Churen Gui}
\author{Shuai Dong}
\email{Corresponding author. Email: sdong@seu.edu.cn}
\affiliation{School of Physics, Southeast University, Nanjing 211189, China}
\begin{abstract}
Nitride perovskites are supposed to exhibit excellent properties as oxide analogues and may even have better performance in specific fields for their more covalent characters. However, till now, very limited nitride perovskites have been reported. In this work, a nitride perovskite LaMoN$_3$ has been systematically studied by first-principles calculations. The most interesting physical property is its ferroelectric $R3c$ phase, which can be stabilized under a moderate hydrostatic pressure ($\sim1.5$ GPa) and probably remain meta-stable under the ambient condition. Its ferroelectric polarization is considerable large, $80.3$ $\mu$C/cm$^2$, driven by the nominal $4d^0$ rule of Mo$^{6+}$, and the covalent hybridization between Mo's $4d$ and N's $2p$ orbitals is very strong. Our calculation not only predicts a new ferroelectric material with prominent properties, but also encourages more studies on pressure engineering of functional nitrides.
\end{abstract}
\maketitle

\textit{Introduction.} Oxides are important quantum materials, which host many exotic emergent physical properties like high-temperature superconductivity, colossal magnetoresistance, multiferroicity, etc. Thus, these oxide families have been extensively investigated in the past decades. Comparing with these widely studied oxides, nitrides have been much less explored. The crucial reasons are their high formation enthalpies and the low chemical potential of N$_2$ molecule \cite{Zakutayev2016-ref1}, which make stable nitrides not as common as oxides.

Even though, nitrides remain highly valuable for their physical properties and promising applications. For example, the blue light-emitting diodes (LED) were based on nitrides  \cite{Nobel-Akasaki2015,Nobel-Amano2015,Nobel-Nakamura2015}, and nowadays devices based on GaN and related materials have been commercially applied widely, which boost the further researches on nitrides.

Very recently, a stability map of inorganic ternary metal nitrides was constructed using the data-mined structure prediction algorithm \cite{Sun2019-ref2}, which provided a useful guide for further explorations of nitrides. Perovskite with the chemical formula $ABX_3$ is the most common crystal structure for ternary metal oxides, and holds a variety of excellent properties for applications. However, the oxidation state of $-3$ for nitrogen requires very high oxidation states for cations at A-site and B-site. Even though, a few nitride perovskites, e.g. LaWN$_3$ and LaMoN$_3$, have been predicted theoretically \cite{Sarmiento-Perez2015-ref3,ref4-Korbel2016}, and ThTaN$_3$ has been synthesized yet \cite{Brese1995}. More oxynitride perovskites, e.g., SrTaO$_2$N, LaTaON$_2$, and LaWO$_\delta$N$_{3-\delta}$, have been experimentally synthesized \cite{ref5-Fuertes2012,ref6-Talley2019,ref7-Clark2013,ref8-Hinuma2012}.

Among these nitride perovskites, LaWN$_3$ was predicted to be ferroelectric with a spontaneous polarization $\sim66$ $\mu$C/cm$^2$ and a small band gap (LDA result: $0.81$ eV) \cite{ref9-Fang2017}. Such a small band gap, due to the spatial-extending $5d$ orbitals, will be an obstruction for experimental verification of its ferroelectricity. In fact, most applied ferroelectric perovskites are based on $3d$ metal oxides, e.g., BaTiO$_3$ and Pb(Zr,Ti)O$_3$, whose $d$ orbitals are more localized, leading to  larger band gaps. Thus, the continuous search for more ferroelectric nitride perovskites remains crucial, especially for those with larger band gaps. A natural idea is to replace the B-site W$^{6+}$ cation using $3d$ or $4d$ cations. However, unluckily, the structure of LaMoN$_3$ was predicted to be non-polar and non-perovskite $C2/c$ phase \cite{Sarmiento-Perez2015-ref3,ref10-Singh2018}.

In this work, based on the first-principles calculations, we predict a pressure-induced structural transition of LaMoN$_3$, from the non-polar non-perovskite $C2/c$ phase to the ferroelectric perovskite $R3c$ phase. And the ferroelectric properties of LaMoN$_3$ can be superior to those of LaWN$_3$.

\textit{Methods.} Density functional theory (DFT) calculations are performed based on the projector augmented wave (PAW) pseudopotentials implemented in Vienna {\it ab initio} simulation package (VASP) \cite{ref11-Kresse1996,ref12-Blochl1994}. The plane wave cutoff energy is fixed to $500$ eV. A $7\times7\times7$ Monkhorst-Pack $\Gamma$-centered $k$-point mesh has been used for structural relax and the Hellmann-Feynman force convergent criterion is set to $10^{-3}$ eV/\AA. The ferroelectric polarization is calculated using the Berry phase method \cite{ref14-King-Smith1993,ref15-Resta1994}, and compared with the point-charge model. To acquire accurate band gaps, the hybrid functional calculation is  performed based on Heyd-Scuseria-Ernzerhof (HSE06) method \cite{Heyd2003,Heyd2004,Heyd2006}.

The positive hydrostatic pressure is applied on LaMoN$_3$ with the stress tensor smaller than $0.1$ GPa throughout our calculations. The enthalpy, which equals the Gibbs free energy at zero temperature, is used to identify the ground state. The dynamic stability of certain structures is verified by the vibrational properties using the density functional perturbation theory (DFPT) \cite{ref13-Gonze1997}. Phonopy is adopted to calculate the phonon band structures  \cite{ref16-Togo2015}, and the AFLOW is used to seek and visualize the dispersion paths in Brillouin zone \cite{ref17-Curtarolo2012}.

\begin{figure}
\centering
\includegraphics[width=0.48\textwidth]{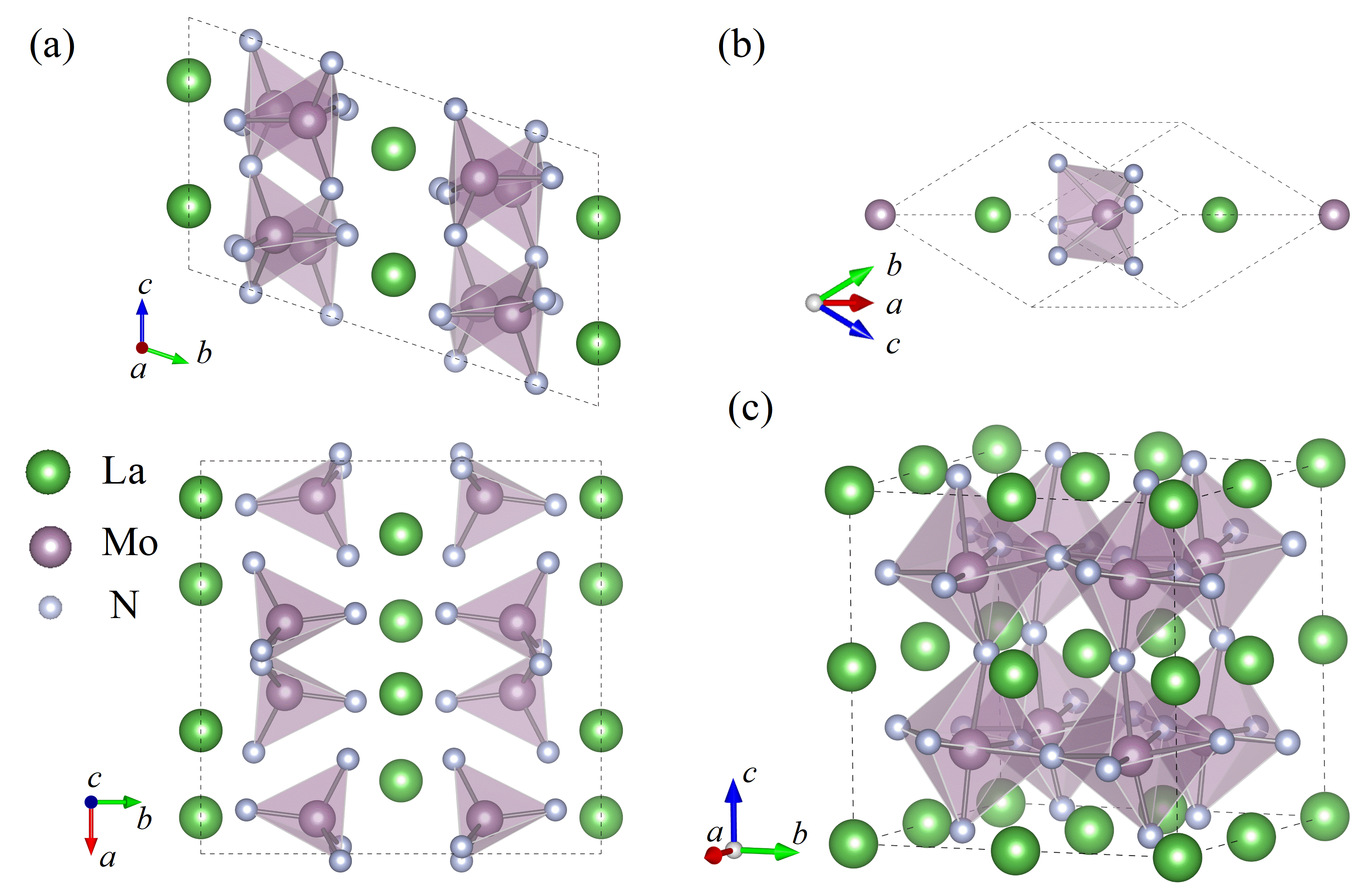}
\caption{Crystal structures of LaMoN$_3$ in (a) the $C2/c$ phase and (b) the $R3c$ phase. (c) The corresponding pseudo-cubic cell of $R3c$ phase.}
\label{fig1}
\end{figure}

\textit{Enthalpy driven structural transition.} As sketched in Fig.~\ref{fig1}(a), the structure of $C2/c$ phase is quite loose, about $26\%$ larger than the compact perovskite one [Fig.~\ref{fig1}(b-c)] \cite{ref10-Singh2018}. The energy difference between the $C2/c$ phase and perovskite $R3c$ phase is only $180$ meV/f.u. \cite{ref10-Singh2018}. Inspired by recent progress of pressure-driven phase transitions in oxides \cite{ref20-Su2019}, we naturally expect a high possibility of phase transition by applying external pressure to LaMoN$_3$.

Referring to the Materials Project database \cite{ref21-Jain2013} and the previous study on LaWN$_3$ \cite{ref9-Fang2017}, totally seven possible structures of LaMoN$_3$ are considered in our calculations: including the pervoskites $R3c$, $R\bar3c$, $Pna2_1$, $P4mm$, and non-perovskite $C2/c$, $Ama2$, $P2_1/m$.

To validate the correctness of our calculation, we first optimized the structure of $C2/c$ phase. Our optimized lattice constants reach good agreements with previous studies [listed in Table~S1 of Supplementary Materials (SM)] \cite{Supp}. Then various phases have been optimized and their energies are compared. Indeed, the $C2/c$ phase has the lowest energy among all these structures, in agreement with the previous structural predictions \cite{Sarmiento-Perez2015-ref3}.

\begin{figure}
\centering
\includegraphics[width=0.4\textwidth]{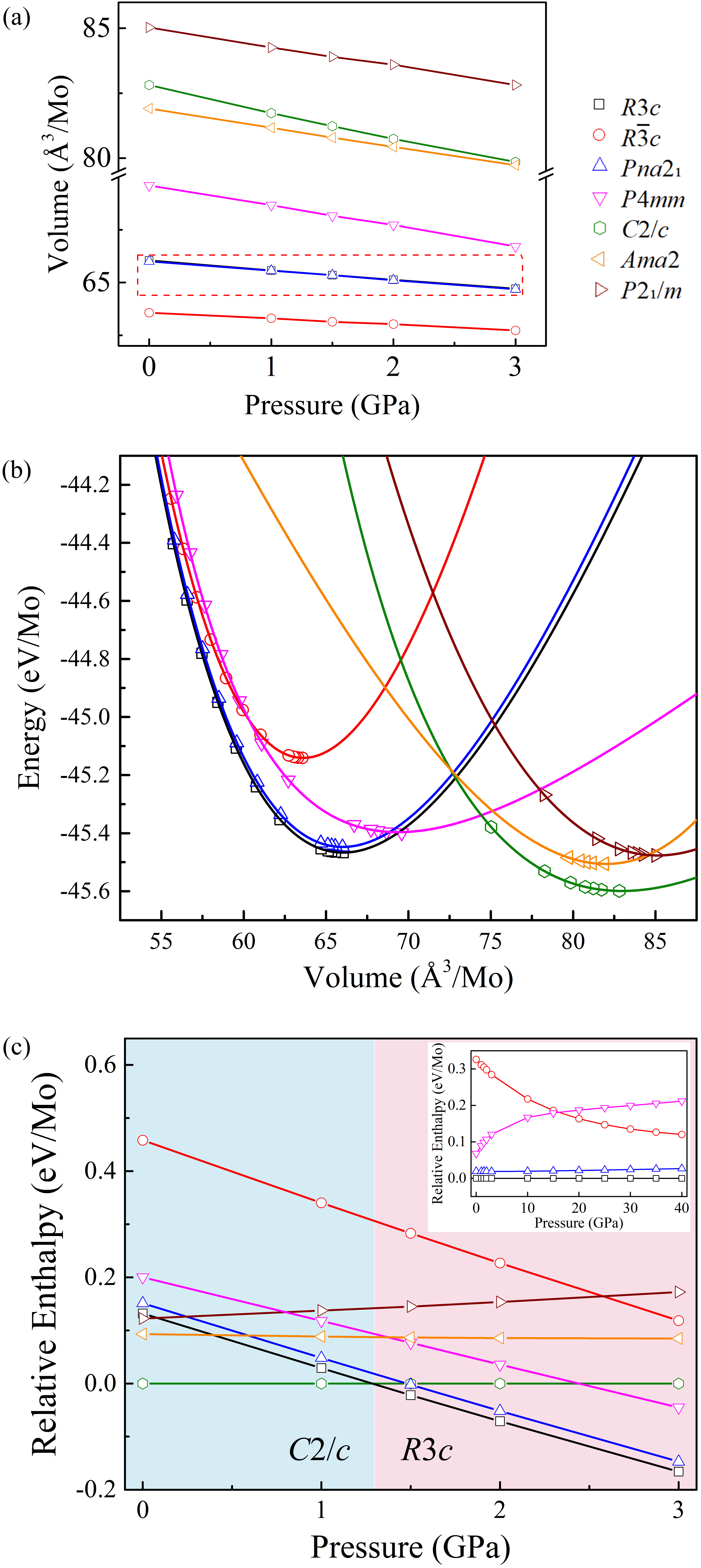}
\caption{(a) The volumes per formula as a function of pressure. The lines of $R3c$ and $Pna2_1$ almost overlap (highlighted in the red-dashed square). At $0$ GPa, the volume of $C2/c$ phase is about $\sim25\%$ larger than the $R3c$ one. (b) The energies of different phases as a function of volume under pressure. The curves are fitted by the Murnaghan equation of state \cite{ref22-Murnaghan1944}. (c) The enthalpies of different phases as a function of pressure. The $C2/c$ phase is taken as the reference (i.e., $0$). The $R3c$ phase becomes more stable from $\sim1.5$ GPa.}
\label{fig2}
\end{figure}

Next, the hydrostatic pressure is applied to all phases. The volumes per formula as a function of pressure are plotted in Fig.~\ref{fig2}(a). In general, the pressure compresses the lattices, and different phases show distinct stiffnesses. Naturally, those loose structures (e.g. the $C2/c$ one) are softer upon pressure. The energy {\it vs} volume curves are shown in Fig.~\ref{fig2}(b).

The enthalpies under pressure are shown in Fig.~\ref{fig2}(c), with reference to the $C2/c$ phase. In principle, the lowest enthalpy value at a given pressure indicates the most stable structure. Thus, a phase transition from the non-polar $C2/c$ phase to the polar $R3c$ phase is expected to occur between $1-1.5$ GPa, a moderate pressure easy to reach in experiment. It is under expectation since the compact perovskite is more favorable  under pressure than the loose $C2/c$ one. Higher pressure up to $40$ GPa region is also tested, and the $R3c$ phase always owns the lowest enthalpy, as shown in the insert of Fig.~\ref{fig2}(c). In other words, no other structural transition exists in this region.

Since the $C2/c$ phase is non-perovskite which can not continuously change to the perovskite one, the phase transition between $C2/c$ and $R3c$ must be the first order one. Thus the $R3c$ phase,  synthesized under pressure, may be meta-stable at the ambient condition after the formation.

More structural details of the $C2/c$ phase at $0$ GPa and $R3c$ phase at $0$/$3$ GPa can be found in Table S2 in SM \cite{Supp}.

\begin{figure}
\centering
\includegraphics[width=0.46\textwidth]{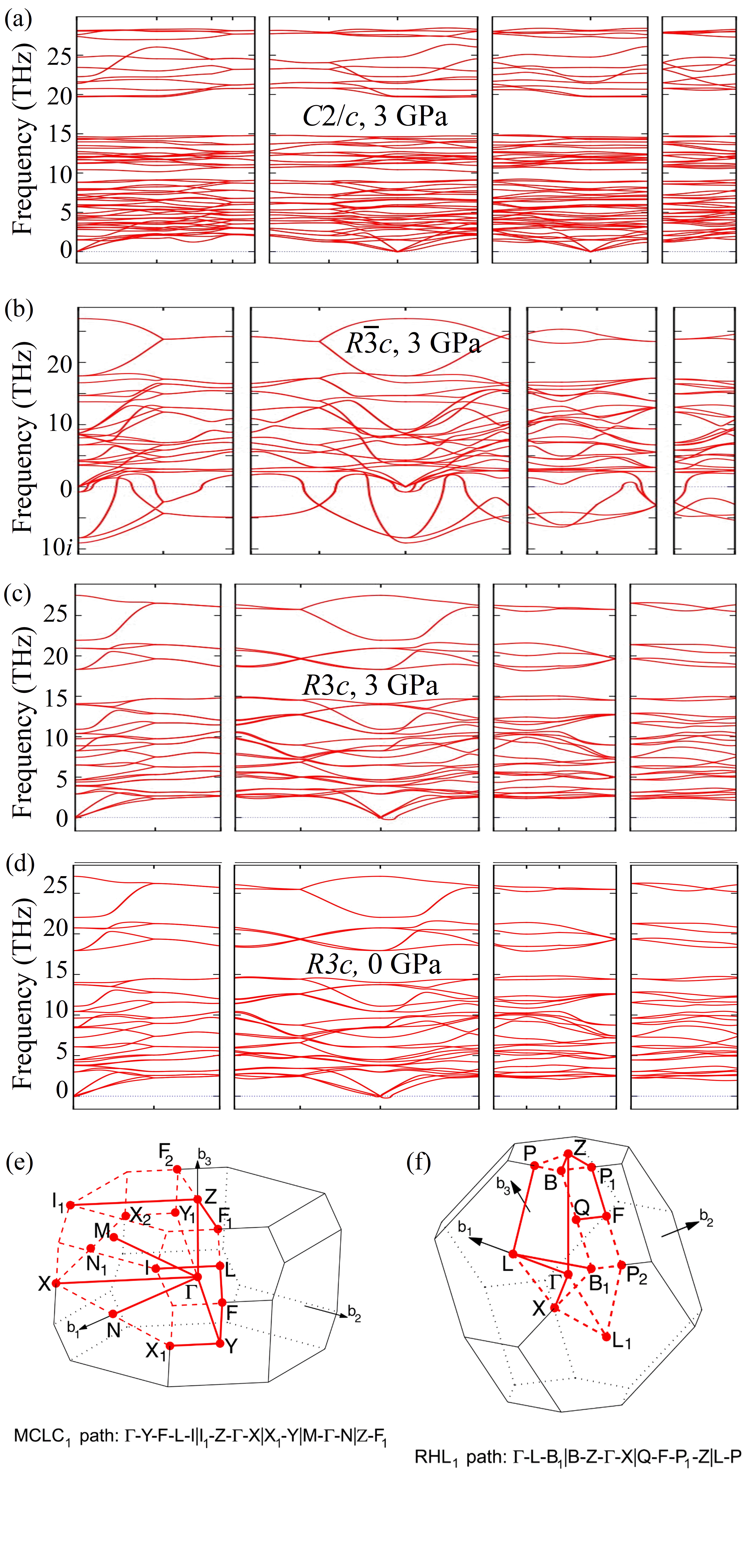}
\caption{Phonon spectra for LaMoN$_3$. (a) The $C2/c$ phase at $3$ GPa, which is meta-stable. (b) The $R\bar3c$ phase at $3$ GPa, which is unstable. (c-d) The $R3c$ one at $3$ GPa and $0$ GPa respectively. %Noting that the tiny imaginary frequencies at non-highly symmetric points are due to numerical inaccuracy of finite size cells used in DFPT calculation, not real physical unstability.
(e-f) The corresponding dispersion pathes between high-symmetric points of Brillouin zone for the $C2/c$ and rhombohedral ($R3c$ and $R\bar3c$), suggested by AFLOW \cite{ref17-Curtarolo2012}.}
\label{fig3}
\end{figure}

\textit{Dynamic stability.} The stability of a structure not only depends on energy/enthalpy, but also needs the dynamic stability. The existence of imaginary vibration mode(s) in the phonon band structure indicates dynamic unstability. Here the dynamic properties of LaMoN$_3$ structures are studied using the DFPT calculations.

First, the phonon spectra of $C2/c$ at $0$ GPa and $3$ GPa have been calculated, as shown in Fig.~S1 of SM and Fig.~\ref{fig3}(a), respectively. No imaginary vibration mode exists in both spectra, indicating the dynamic stability. In particular, the dynamic stability of $C2/c$ phase at $3$ GPa implies its meta-stable fact under pressure, a character of first order phase transition.

Next, the $R\bar3c$ phase is considered, which is the parent group of $R3c$, i.e. the paraelectric state. As shown in Fig.~\ref{fig3}(b), some unstable phonon modes appear around the $\Gamma$ point, which is an evidence of ferroelectric transition. Following the eigenvector of the most unstable phonon mode, i.e., the one with the highest imaginary frequency,  the $R3c$ phase with lower symmetry is obtained. Then the phonon bands of $R3c$ at $3$ GPa  have been calculated, which show dynamic stability [Fig.~\ref{fig3}(c)]. Furthermore, after removing the pressure, no serious imaginary mode exists in the phonon spectrum of $R3c$ one at $0$ GPa, implying the meta-stable fact of this ferroelectric phase at the ambient condition.

\begin{figure}
\centering
\includegraphics[width=0.48\textwidth]{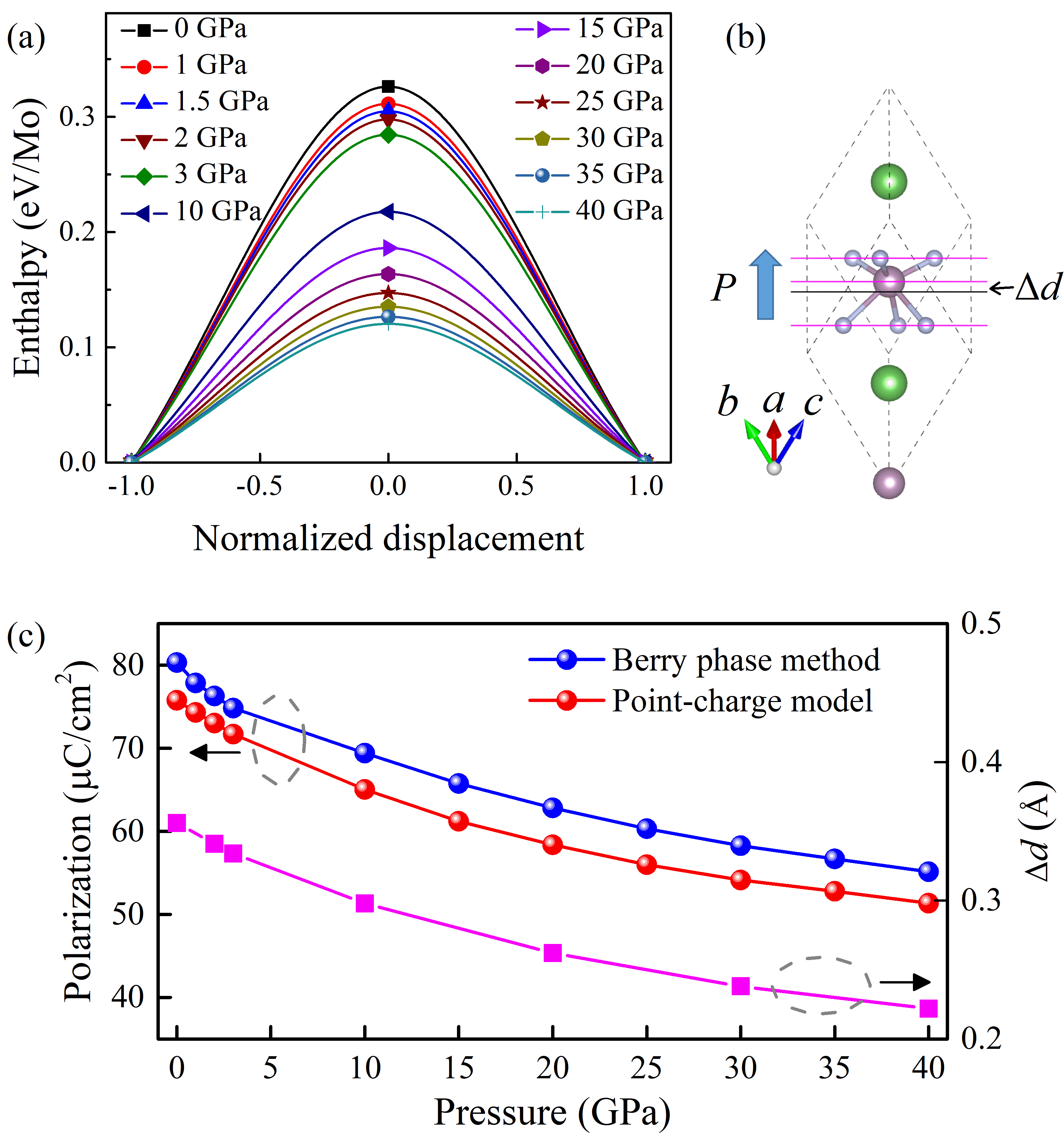}
\caption{(a) The energy difference between $R3c$ and $R\bar3c$ as the barrier for polarization switching under different pressures. (b) The ferroelectric displacement (defined as $\Delta d$) of Mo ion along the [111] direction. Magenta lines: the (111) planes for Mo and N$_3$ ions; Black line: the original mirror plane of the paraelectric state. (c) Polarization (left) and ferroelectric displacement (right) as a function of pressure. The polarization calculated by Berry phase method is accurate, which is slightly larger then the intuitional one estimated by the point-charge model. The ferroelectric displacement is suppressed by pressure, which is the reason for the reduced polarization and ferroelectric switching barrier.}
\label{fig4}
\end{figure}

\textit{Ferroelectricity \& electronic structure.} To explore the ferroelectricity of $R3c$ phase, the spontaneous polarization and ferroelectric switching barrier are calculated, as shown in Fig.~\ref{fig4}(a). At $0$ GPa, the polarization estimated by the Berry phase method is $80.3$ $\mu$C/cm$^2$, larger than that of LaWN$_3$ ($66$ $\mu$C/cm$^2$ \cite{ref9-Fang2017}) and comparable to PbTiO$_3$ (a famous large-polarization ferroelectric oxide whose polarization reaches $\sim80$ $\mu$C/cm$^2$ \cite{Bonini2020}).

By analyzing the structural distortion, it is clear that such ferroelectric polarization originates from the displacement of Mo$^{6+}$ ion towards the diagonal direction of N$_6$ octahedron, as shown in Fig.~\ref{fig4}(b). This mechanism is called the $d^0$ rule, according to the experience of ferroelectric perovskite oxides. However, in typical ferroelectric perovskite oxides, e.g. BaTiO$_3$, the $d^0$ rule usually drives a ferroelectric distortion toward one neighbor oxygen anions, instead of the diagonal direction of O$_6$ octahedron.

The ferroelectric structure of LaMoN$_3$ is somewhat similar to BiFeO$_3$ which also owns the $R3c$ structure \cite{Hatt2010}. However, the ferroelectric polarization in BiFeO$_3$ is driven by the $6s^2$ lone pair of Bi$^{3+}$, not the B-site cation. Anyhow, the plenty experience of ferroelectricity from BiFeO$_3$ can be applied to LaMoN$_3$. For example, for such $R3c$ perovskite, the polarization is along the eight-fold $<111>$ directions of pseudo-cubic cell, which leads to the possible $71^\circ$, $109^\circ$, and $180^\circ$ domain walls \cite{Seidel2009}.

The polarization is suppressed moderately by pressure, as shown in Fig.~\ref{fig4}(c), e.g. $\sim74.8$ $\mu$C/cm$^2$ at $3$ GPa. The energy barrier, i.e. the energy/enthalpy difference between the ferroelectric and paraelectric state is about $0.3$ eV/Mo (the upper limit in the real flop-over process while another lower energy pathes may be possible), which is also suppressed continuously by pressure [Fig.~\ref{fig4}(a)].

Another interesting result is that the polarization estimated using the point-charge model is slightly lower than the Berry phase one, as compared in Fig.~\ref{fig4}(c). This is not common  but suggests the deviation from ideal ionic crystal. Due to the weak electronegativity of N and high valences of N/Mo, there must be a lot of electrons distributing in the partially covalent N-Mo bonds, instead of staying in N and Mo sites locally. Then the deformation of electron cloud can contribute a lot to the total polarization, even the ions do not move too much.

\begin{figure}
\centering
\includegraphics[width=0.46\textwidth]{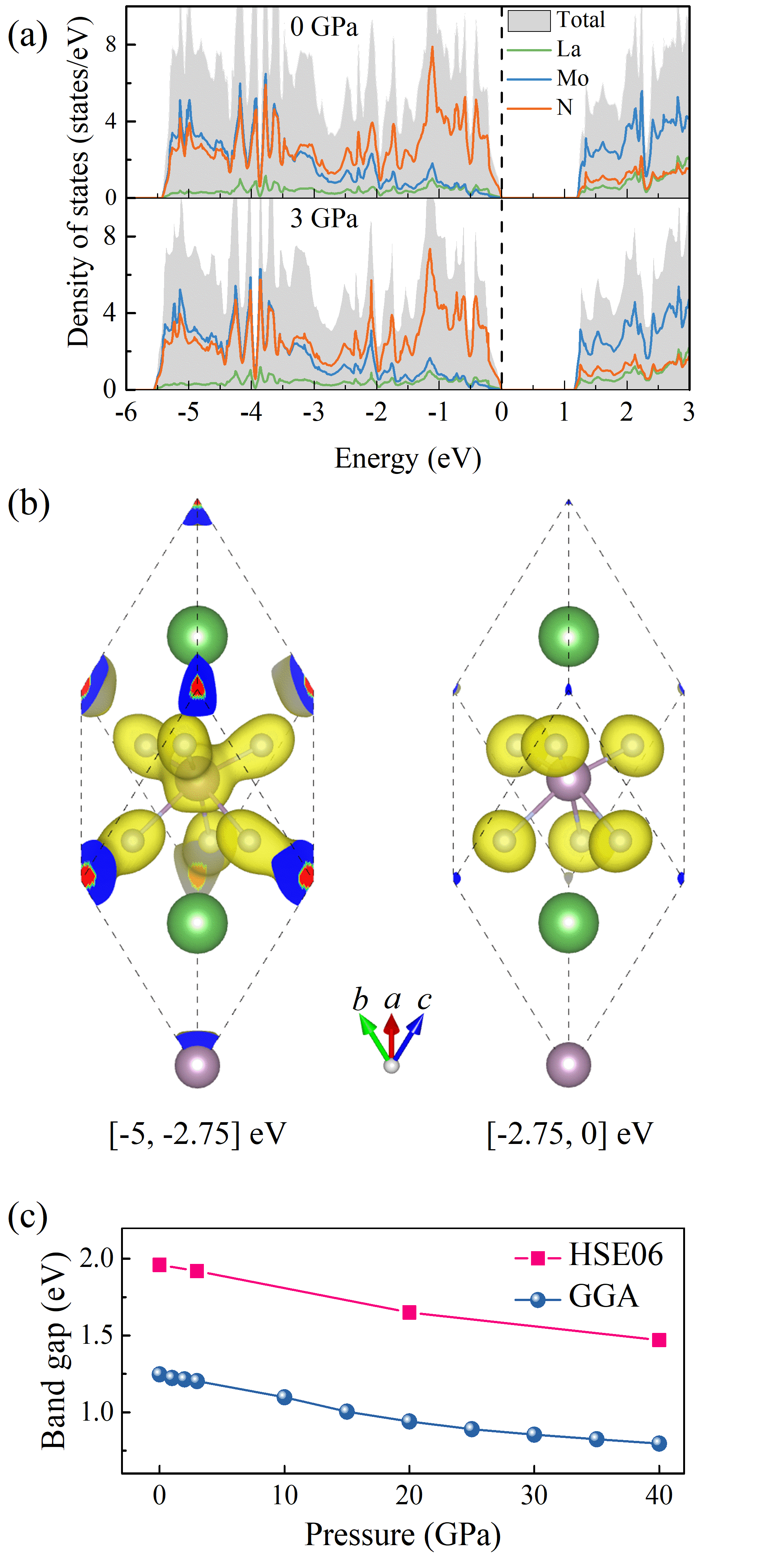}
\caption{(a) The electronic DOS under $0$ GPa and $3$ GPa. The Fermi level is set at $0$ eV. The strong hybridization between N's and Mo's orbitals occurs mainly in the [$-5.5$, $-2.75$] eV region. (b) The spatial electron distributions for the [$-5.5$, $-2.75$] eV and [$-2.75$, $0$] eV energy windows. The strong covalent Mo-N bonds can be evidenced in the [$-5.5$, $-2.75$] eV case. (c) The GGA and HSE06 band gaps as a function of pressure.}
\label{fig5}
\end{figure}

The partially covalent characteristic of Mo-N bonds can also be evidenced in the electronic density of states (DOS), as shown in Fig.~\ref{fig5}(a). There is strong hybridization between N's $2p$ and Mo's $4d$ orbitals, although nominally the $4d$ orbitals should be empty in the ionic crystal limit. Such hybridization and covalent characteristic of Mo-N bonds can be also visualized by the electron spatial distribution, as shown in Fig.~\ref{fig5}(b).

According to Fig.~\ref{fig5}(a), the pressure can slightly reduce the band gap, which can be understood as the enhanced bandwidth of more compact lattice. The band gap of $R3c$ phase is plotted in Fig. 5(c) as a function of pressure. Although the band gap at $0$ GPa is only $1.25$ eV for the GGA calculation, the hybrid functional calculation based on HSE06 gives $1.96$ eV, which is certainly larger than that of LaWN$_3$ ($0.81$ eV for the LDA calculation and $1.72$ eV for the HSE06 calculation \cite{ref9-Fang2017}).

Last, the magnetism is also tested considering the fact that Mo's $4d$ orbitals are partially ``occupied" in the shared manner. However, no net magnetic moment is found in our GGA calculation, even under pressure up to $40$ GPa. Thus the $R3c$ phase of LaMoN$_3$ is pure ferroelectric, not multiferroic.

\textit{Conclusion.} In summary, the pressure-induced structural phase transition from the non-perovskite non-polar $C2/c$ phase to the polar perovskite $R3c$ phase in LaMoN$_3$ has been studied using first-principles calculations. Such a transition occurs at a moderate pressure less than $1.5$ GPa, which is accessible in experiments. This polar structure remains the ground state up to high pressures, and is probably meta-stable at the ambient condition. The ferroelectric properties of LaMoN$_3$ is prominent, superior to LaWN$_3$. The partially covalent bond between Mo and N ions are responsible for the ferroelectricity. Further experimental studies are encouraged to verify our prediction and find more functional nitrides.

\begin{acknowledgments}
This work was supported by the National Natural Science Foundation of China (Grant Nos. 11834002 and 11674055). We thank the Tianhe-II of the National Supercomputer Center in Guangzhou (NSCC-GZ) and the Big Data Center of Southeast University for providing the facility support on the numerical calculations.
\end{acknowledgments}

\bibliographystyle{apsrev4-1}
\bibliography{LaMoN3}
\end{document}